**Quadrilaterals on the square screen of their diagonals: Regge symmetries of quantum-mechanical spin-networks and Grashof classical mechanisms of four-bar linkages.**

Vincenzo Aquilanti [1,2], Ana Carla Peixoto Bitencourt [3], Concetta Caglioti [1], Robenilson Ferreira dos Santos [1,4,5], Andrea Lombardi [1], Federico Palazzetti [1,*], Mirco Ragni [3].

[1] Dipartimento di Chimica, Biologia e Biotecnologie, Università di Perugia, via Elce di Sotto 8, 06123 Perugia, Italy.

[2] Istituto di Struttura della Materia - Consiglio Nazionale delle Ricerche, 00016 Rome, Italy.

[3] Departamento de Física, Universidade Estadual de Feira de Santana, Avenida Transnordestina s/n, 44036-900 Feira de Santana, BA, Brazil.

[4] Instituto de Fìsica, Universidade Federal da Bahia, Campus Universitario de Ondina, CEP 40210-340 Salvador, BA, Brazil.

[5] Instituto Federal de Alagoas - Campus Piranhas, CEP 57460-000 Piranhas, AL, Brazil.

* Corresponding author: federico.palazzetti@unipg.it

*Abstract*

The four-bar linkage is a basic arrangement of mechanical engineering and represents the simplest movable system formed by a closed sequence of bar-shaped bodies. Although the mechanism can have in general a spatial arrangement, we focus here on the prototypical planar case, starting however from a spatial viewpoint. The classification of the mechanism relies on the angular range spanned by the rotational motion of the bars allowed by the ratios among their lengths and is established by conditions for the existence of either one or more bars allowed to move as cranks, namely to be permitted to rotate the full 360° range (Grashof cases), or as rockers with limited angular ranges (non-Grashof cases). In this paper, we provide a view on the connections between the "classic" four-bar problem and the theory of $6j$ symbols of quantum mechanical angular momentum theory, occurring in a variety of contexts in pure and applied quantum mechanics. The general case and a series of symmetric configurations are illustrated, by representing the range of existence of the related quadrilaterals on a square "screen" (namely as a function of their diagonals) and by discussing their behavior according both to the Grashof conditions and to the Regge symmetries, concertedly considering the classification of the two mechanisms and that of the corresponding objects of the quantum mechanical theory of angular momentum. An interesting topological difference is demonstrated between mechanisms belonging to the two Regge symmetric configurations: the movements in the Grashof cases span chirality preserving configurations with a $2\pi$-cycle of a rotating bar, while by contrast the non-Grashof cases span both enantiomeric configurations with a $4\pi$-cycle.

## 1. Introduction.

The four-bar linkage (see for example Hartenberg *et al.* 1964 and Toussaint 2003) was introduced during the Industrial Revolution and since then is a pervasively common mechanism occurring in wide area that spans from traditional engineering and robotics to even biomechanics, biochemistry, microbiology, and molecular dynamics. Generally, the four bars can be assembled in three dimensions, although the most common configurations concern planar displacements; this is the case upon which we focus in this paper (Figure 1). The ratios between the lengths of the bars determine different paths of the joints that permit to classify the performances of the linkage. One of the four bars is usually fixed, the *ground link*, and is directly connected to an *input link* and an *output link*. The remaining bar, the one not connected to the ground link, is called the *floating link*. To be specific, we consider a rotation angle $\theta$ defined by the direction of the input link with respect to the ground link, conventionally drawn as horizontal, as in Figure 1. The input link movements are classified according to four possible kinds of allowed rotations: the *crank* – the bar rotates by the full $\theta = 360°$ range; the *rocker* – the rotation is allowed within a limited range that does not include $\theta = 0°$ and $180°$; the 0-*rocker* – where different from the *rocker*, the rotation excludes $\theta = 180°$; the $\pi$-*rocker,* where differently from the rocker and from the previous case, the rotation excludes $\theta = 0°$. Four kinds of rotations can be immediately individuated according to the Grashof conditions by comparing the lengths of the four-bars: they are said to accomplish the Grashof *condition* given by equation (1), this takes place when the shortest (s) of the four bars can rotate fully with respect to the neighboring bars, if the sum of the longest (l) and the shortest bars is less than or equal to the sum of the remaining two (p and q) (Fig. 1a):

$$s + l \leq p + q. \qquad (1)$$

We will see later that four-bar linkages obeying to the Grashof condition belong to eight cases, according to the positive or negative sign of appropriate combinations of the lengths, based on analysis of angle existence of input link (Muller 1996, Mc Carthy *et al* 2011, Mauroy *et al* 2016).

In this work, we reconsider this classification exploiting similarity with features encountered in case studies of the coupling of angular momentum (Ragni *et al.* 2013): to shed further light on the four-bar mechanism problem, we establish connections between the conditions of Grashof and the symmetries of Regge of the 6*j* symbols, through the *screen* representation (Aquilanti *et al.* 2008a, Bitencourt *et al.* 2012, 2014). The screen originates from the geometrical interpretation of the 6*j* symbols in the semiclassical limit, through the Ponzano-Regge theory (Ponzano *et al.* 1968), the application of the properties of tetrahedra and quadrangles, and the symmetry properties of the 6*j* symbols (Anderson *et al.* 2006, 2008, 2009). It was initially developed for the representation of the Regge symmetries of 6*j* symbols, the building blocks of spin networks (Anderson *et al.* 2013, Anderson 2014, Anderson *et al.* 2017, Aquilanti *et al.* 2009, 2018, Arruda *et al.* 2016), and then applied to compact the available data of peroxides (Maciel *et al.* 2006, 2007a, 2007b, Barreto *et al.* 2007) and persulphides (Maciel *et al.* 2008, by displaying the distances (Aquilanti *et al.* 2017a) related to chirality changes (Moss 1996, Aquilanti *et al.* 2008a, Lombardi *et al.*2017).

The paper is structured as follows: in Section 2, we give background information related to the quadrilaterals, quadrangles and tetrahedra, the formulas adopted to build the screen and its features, the symmetry relations of 6*j* symbols, including the symmetry of Regge and give a formal enunciation of the Grashof classification condition; in Section 3, we discuss features from the viewpoints of the screen for the general case and for symmetric or degenerate ones; in Section 4, concluding remarks end the paper.

## 2. Background.

### 2.1. Quadrilaterals, quadrangles and tetrahedra.

A quadrilateral is a polygon, that in planar Euclidean geometry is defined as a geometric shape with four sides, the sum of the four inner angles being 360°. The opposite vertices of quadrilaterals are joined by diagonals. Their classification relies on the displacement of the diagonals: convex (two inner diagonals); biconcave (two outer diagonals); concave (one inner and one outer diagonal). In projective and affine geometries, a complete quadrilateral is formed by four points and six lines, being diagonals considered as further sides. A complete quadrangle is a planar configuration of four incident lines in six points. In projective geometry, quadrangles and quadrilaterals are the same object, since duality in the real projective plane $RP_2$ permits exchange of the role of points and lines. In space, objects of non-zero volume are seen as tetrahedral, being formed by four vertices and six edges.

In the four-bar mechanism, the four links (see Sec. 1) and the diagonals connecting the opposite joins can thus be seen as the edges of a tetrahedron, where the planar displacement represents those of a tetrahedron of zero volume. The length of the four bars and of the two diagonals represent a 6-distance system that shows analogies with the 6*j* symbols of Wigner, Racah coefficients. Wigner and Racah gave a geometrical interpretation of these elements, attributing to the six entries of the 6*j* as the edges of a general irregular tetrahedron. (Ponzano *et al.* 1968, Neville 1971, 2006, Schulten *et al.* 1975a, 1975b, Ragni *et al.* 2010, Aquilanti *et al.* 2000, 2001a).

## 2.2. The screen.

A 4$j$ model (Littlejohn *et al.* 2009, Aquilanti *et al.* 2012) permits to represent the allowed range of a tetrahedron through a two-dimensional plot of a couple of discrete variables for the 6$j$ symbol (Aquilanti *et al.* 2008b, 2013, 2017b)), which entries are designates as $j_1, j_2, j_3, j_4, j_{12}, j_{23}$ (see also Aquilanti *et al.* 2001b, 2007, Santos *et al.* 2017). An analogous 4$d$ model (Aquilanti *et al.* 2017a) is introduced here to represent the six distances and the geometrical properties of the associated tetrahedron described in Section 2.1. According to the "canonical" ordering (Bitencourt *et al.* 2014), we denote the edges as follows: $a$ is the shortest one, $c$ is opposite to $a$ and $d$ is the longest one between the remaining two edges. These four edges are fixed by construction and in the quadrangle (the tetrahedron projected on the plane) they are its sides. The remaining two edges, namely $x$ and $y$, are variables and for the planar case they are the diagonals of the quadrilateral. The resulting "screen" on the plane, defined by the allowed ranges $x_{min}$, $x_{max}$, $y_{min}$, $y_{max}$, is a square.

$$x_{min} = b - a$$

$$x_{max} = b + a \qquad (2)$$

$$y_{min} = d - a$$

$$y_{max} = d + a.$$

Sometimes it is expedient to confine the screen into a square, which sides range between 0 and 1, one has to reduce the distances corresponding to $x$ and $y$, respectively, by $b - a$ and $d - a$ and normalized by $b + a$ and $d + a$.

In the screen, we define two significant curves called *caustic* and *ridge*. The caustic delimit the existence of a quadrilateral and surrounds an area corresponding to a tetrahedron (the caustic corresponds to a tetrahedron of zero volume). This line touches four points of the axes of the screen, called *gates*, indicated by the four cardinal points north N, south (S), east (E) and west (W). The north gate identifies the configuration for which the value of $x$ corresponds to $y$ = 1, while S is the counterpart for $y$=0. Analogously, E and W are the values of $y$ corresponding to $x$=0 and $x$=1, respectively. The *ridges* are curves that mark configurations of the associate tetrahedron when two specific pairs of triangular faces are orthogonal.

<u>Caustic curve.</u> The equation of the caustic has been derived by the Heron formula that permits to calculate the area of a triangle as a function of the sides. The quadrangle can be seen as two triangles with a common side: *abx* and *cdx*, whose areas are $F^x{}_{a,b}$ and $F^x{}_{c,d}$.

$$F_{a,b}^x = \left\{ \frac{[(a+b-x)(-a+b+x)(a-b+x)(a+b-x)]}{16} \right\}^{1/2} \quad (3)$$

$$F_{a,b}^x = \left\{ \frac{[((a+b)-x)((-a+b)+x)((a-b)+x)((a+b)-x)]}{16} \right\}^{1/2}. (4)$$

Equations (3) and (4) are rearranged to give:

$$F_{a,b}^x = \left\{ \frac{[(a+b)^2 - x^2][x^2 - (b-a)^2]}{16} \right\}^{1/2} \quad (5)$$

$$F_{c,d}^x = \left\{ \frac{[(c+d)^2 - x^2][x^2 - (d-c)^2]}{16} \right\}^{1/2}. (6)$$

Thus, the area of the quadrangle is

$$F_{a,b}^x + F_{c,d}^x = \left\{ \frac{[(a+b)^2-x^2][x^2-(b-a)^2]}{16} \right\}^{1/2} + \left\{ \frac{[(c+d)^2-x^2][x^2-(d-c)^2]}{16} \right\}^{1/2}. (7)$$

The area of the quadrangle, $K$, can also be calculated by the Bretschneider generalized formula (Coolidge 1939):

$$K = \left\{ \frac{[4x^2y^2 - (b^2+d^2-a^2-c^2)^2]}{16} \right\}^{1/2}. (8)$$

Thus

$$K = F_{a,b}^x + F_{c,d}^x. \quad (9)$$

By substituting equation 9 with equations 4 and 5 one obtains:

$$\left\{ \frac{[4x^2y^2 - (b^2+d^2-a^2-c^2)^2]}{16} \right\}^{1/2} = \left\{ \frac{[(a+b)^2 - x^2][x^2 - (b-a)^2]}{16} \right\}^{1/2}$$

$$+ \left\{ \frac{[(c+d)^2 - x^2][x^2 - (d-c)^2]}{16} \right\}^{1/2}. \quad (10)$$

In order to eliminate the square root of both sides, we square the two terms of Equation 9:

$$K^2 = \left( F_{a,b}^x + F_{c,d}^x \right)^2 = \left( F_{a,b}^x \right)^2 + \left( F_{c,d}^x \right)^2 + 2F_{a,b}^x F_{c,d}^x. \quad (11)$$

In such a way, the root terms are eliminated and obtained by a diagonal analytical expression as a side function. By replacing both sides of Equation (11) by Equations (5), (6) and (8), one obtains the following expression:

$$y^2 = \frac{4}{x^2}\left\{\frac{[(b^2-a^2)+(d^2-c^2)]^2}{16} + \left(F_{a,b}^x\right)^2\left(F_{c,d}^x\right)^2 + 2F_{a,b}^x F_{c,d}^x\right\} \quad (12)$$

$$y^2 = \left\{\frac{[(b^2-a^2)+(d^2-c^2)]^2}{x^2} + \frac{4\left[\left(F_{a,b}^x\right)^2 + \left(F_{c,d}^x\right)^2\right]}{x^2} + \frac{8F_{a,b}^x F_{c,d}^x}{x^2}\right\} \quad (13)$$

$$y^2 = \left\{\frac{[(b^2-a^2)+(d^2-c^2)]^2}{x^2} + \frac{4\left[\left(F_{a,b}^x + F_{c,d}^x\right)^2\right]}{x^2}\right\}. \quad (14)$$

Equation 14 is named caustic and relates the diagonals of a quadrilateral with only the respective sides and respective two-sided areas that share the same side. An alternative way consists in obtaining the same expression only as a function of its sides, in this case we substitute Equations (5) and (6) in Equation (12):

$$\frac{4x^2y^2\,(b^2+d^2-a^2-c^2)^2}{16}$$
$$= \frac{[(a+b)^2-x^2][x^2-(b-a)^2]}{16} + \frac{[(c+d)^2-x^2][x^2-(d-c)^2]}{16}$$
$$\pm 2\,\frac{\left[[(a+b)^2-x^2][x^2-(b-a)^2][(c+d)^2-x^2][x^2-(d-c)^2]\right]^{\frac{1}{2}}}{16} \quad (15)$$

obtaining

$$y^2 = \frac{1}{4x^2}\left\{(b^2+d^2-a^2-c^2)^2 + [(a+b)^2-x^2][x^2-(b-a)^2] + [(c+d)^2-x^2][x^2-(d-c)^2] \pm 2\left[[(a+b)^2-x^2][x^2-(b-a)^2][(c+d)^2-x^2][x^2-(d-c)^2]\right]^{\frac{1}{2}}\right\}. \quad (16)$$

This latter equation relates the diagonals and the sides of the quadrilateral. It corresponds to a tetrahedron when its volume is zero. For Ponzano and Regge 1968, the volume of a tetrahedron is given by:

$$V = \frac{2}{3x}F_{a,b}^x\,F_{c,d}^x\sin\theta \quad (17)$$

From Equations (4) and (5), we can formulate Equation 17 as a function of the edges of the tetrahedron:

$$V = \frac{2}{3x}\left\{\frac{[(a+b)^2-x^2][x^2-(b-a)^2]}{16}\right\}^{1/2}\left\{\frac{[(c+d)^2-x^2][x^2-(d-c)^2]}{16}\right\}^{1/2}\sin\theta \quad (18)$$

<u>The ridges.</u> The maximum value of the volume of a tetrahedron is given for $\theta = \pi/2$. From Equation (18), one obtains the equations of the *ridges* that correspond to the maximum volume for the tetrahedron obtained by folding the planar quadrilaterals either on the $x$ or on then $y$ diagonal:

$$y_R^2 = \frac{1}{4x^2}\{ (b^2 + d^2 - a^2 - c^2)^2 + [(a+b)^2 - x^2][x^2 - (b-a)^2] + [(c+d)^2 - x^2][x^2 - (d-c)^2] \} \quad (19)$$

$$x_R^2 = \frac{1}{4y^2}\{ (b^2 + d^2 - a^2 - c^2)^2 + [(d+a)^2 - y^2][y^2 - (d-a)^2] + [(c+b)^2 - y^2][y^2 - (c-b)^2] \} \quad (20)$$

<u>The gates.</u> Finally, the four *gates* are given by the following equations:

for $x = b - a$, corresponding to the W, one has

$$y_W^2 = \frac{1}{4(b-a)^2}\{ [(b-a^2) + (d^2 - c^2)]^2 + [(c+d)^2 - (b-a)^2][(b-a)^2 - (d-c)^2] \} \quad (21)$$

while for $x = b + a$, corresponding to the E:

$$y_E^2 = \frac{1}{4(b+a)^2}\{ [(b^2 - a^2) + (d^2 - c^2)]^2 + [(c+d)^2 - (b+a)^2][(b+a)^2 - (d-c)^2] \}. \quad (22)$$

For $y = d - a$, corresponding to the S:

$$x_S^2 = \frac{1}{4(d-a)^2}\{ [(b^2 - a^2) + (d^2 - c^2)]^2 + [(b+c)^2 - (d-a)^2][(d-a)^2 - (c-b)^2] \} \quad (23)$$

and for $y = d + a$, corresponding to the N:

$$x_N^2 = \frac{1}{4(d+a)^2}\{ [(b^2 - a^2) + (d^2 - c^2)]^2 + [(b+c)^2 - (d+a)^2][(d+a)^2 - (c-b)^2] \}. \quad (24)$$

## 3. Case studies and discussion.

### 3.1 Correspondence between Regge symmetries and Grashof cases.

A four-bar linkage displays different configurations, according to the relative lengths of the bars. As mentioned in the Introduction, the bars are classified as ground, input, output and floating links (McCarthy *et al.* 2011). Systems are classified according to appropriate combinations of the length of the four bars, that also provide information regarding the Grashof condition:

$$T_0 = g + a + h + b \quad (25)$$
$$T_1 = g - a + h - b \quad (26)$$
$$T_2 = g - a - h + b \quad (27)$$
$$T_3 = -g - a + h + b \quad (28)$$

where $g$ is the ground link, $a$ is the input link, $b$ is the output link and $h$ is the floating link. Equation (25) has to be added to build up an orthogonal transformation upon $a, g, b, h$. In Figure 1, we show the comparison between a quadrangle originated by the coupling scheme of angular moments and a similar quadrangle representing a four-bar linkage. In the first case, the sides and diagonals of the quadrangle are represented by the angular momenta, while in the second case they are represented by the length of the four bars and the two distances between the opposite angles of the quadrangle.

The Regge symmetry relations (Section A.1) can be formulated by a set of variables *s, u, v and r* introduced to describe the topology of the corresponding screen:

$$\begin{Bmatrix} a & b & x \\ c & d & y \end{Bmatrix} = \begin{Bmatrix} a' & b' & x \\ c' & d' & y \end{Bmatrix} = \begin{Bmatrix} s-a & s-b & x \\ s-c & s-d & y \end{Bmatrix} = \begin{Bmatrix} c-r & d+r & x \\ a-r & b+r & y \end{Bmatrix} = \begin{Bmatrix} b-u & a-u & x \\ d+u & c+u & y \end{Bmatrix} =$$

$$\begin{Bmatrix} d-v & c+v & x \\ b+v & a-v & y \end{Bmatrix} \tag{29}$$

where the variables are defined as follows:

*s* is the semiperimeter:          $s = \frac{1}{2}(a + b + c + d)$          (30)

*u* is the difference between lines:          $u = \frac{1}{2}[(a + b) - (c + d)]$          (31)

*r* is the difference between columns:          $r = \frac{1}{2}[(a + c) - (b + d)]$          (32)

*v* is the difference between diagonals:          $v = \frac{1}{2}[(a + d) - (b + c)]$          (33)

Comparison between Equations (25) – (28) and (30) – (33) gives the following relations:

$$T_0 = 2s,$$

$$T_1 = 2r;$$

$$T_2 = 2u,$$

$$T_3 = 2v. \tag{34}$$

Table I shows the sign (+, -, 0) of the Regge variables (*r, u* and *v*) for each of the 27 four-bar linkage types, classified according to the input and output links. Systems I – I', II- II', III – III' and IV – IV' are couples of Regge conjugates, for all of them the value of *r, u* and *v* is different from zero. Systems labelled from 1 to 19 are instead characterized by at least one of the Regge variables equal to 0.

### 3.2. General case: illustrations.

In this Section, we discuss the screen (Figure 2) for a system characterized by the sides: $a = 30$, $b = 45$, $c = 60$, $d = 55$. It belongs to the case IV', being $u = -20$, $r = -5$ and $v = -10$. The range of the diagonals is $15 \le x \le 75$ and $25 \le y \le 85$

$$\begin{Bmatrix} a & b & x \\ c & d & y \end{Bmatrix} = \begin{Bmatrix} 30 & 45 & x \\ 60 & 55 & y \end{Bmatrix} = \begin{Bmatrix} a' & b' & x \\ c' & d' & y \end{Bmatrix} = \begin{Bmatrix} 65 & 50 & x \\ 35 & 40 & y \end{Bmatrix} \tag{35}$$

Since the caustic does not present symmetry features, the case is indicate as general. Figures 3 and 4 report the configurations of the four-bars linkage in correspondence of the four gates N, W, E, S and in the intermediate points NW, NE, SW, SE.

Figure 3 shows the four-bar linkage, for which the Grashof condition is verified. Diagonals $x$ and $y$ vary according to the caustic reported in Figure 1. The side $b$ is considered the ground link, while the side $a$, the input or output link, works as a crank, being allowed to rotate of 360° clockwise (left column) or anticlockwise (right column). The first configuration we report (left column) consists of the sides $a$ and $b$ aligned and the $x$ diagonal is at its maximum length; it corresponds to the E gate. The evolution of the system leads to concave quadrilaterals SE, then in S it assumes the lowest value of the $y$ diagonal, with $a$ lying over $d$. The passage to the gate SW leads to a biconcave quadrilateral and subsequently to W, where $x$ reaches the minimum length, with $a$ folded over $b$. A further rotation to NW, concave quadrilateral, and N yields the highest length of $y$, with $a$ and $d$ aligned, and finally a convex quadrilateral in correspondence of NE. After a 360° rotation of the side $a$ in correspondence of E, we obtain a configuration that is specular to the initial one (compare the E configurations of the first and second column). By further rotating the $a$ side, one obtains configurations specular to those reported in the first column (see the second column), until reaching the E configuration in the first column after a second 360° rotation. In Figure 4, we illustrate the four-bar linkage for the Regge conjugate of the object shown in Figure 3. In Table IV, the relation transformations of the Regge symmetry are reported. The Regge symmetry produces an object for which the Grashof condition is not accomplished. In this case, the bar $b$' (left column) moves towards the ground link $a$'. As seen previously for the case of Figure 3, in correspondence of the gates two bars are aligned, forming a triangle: the diagonals ($x$ for E and $y$ for N) reach the maximum length corresponding to $c$'+$d$' and $c$'+$b$', respectively. At W and S the diagonals ($x$ for W and for at S) assume the minimum value, given by $a$'-b' and $a$'-$d$', respectively. In correspondence of the intermediate points SE, SW, NW and NE, the configurations are those of a concave, biconcave, concave, and convex quadrilateral, respectively. Being a non-Grashof case, $b$' does not work as a crank, but inverts the direction of rotation. From a topological point of view, the sequence of configurations for both the Grashof and the non-Grashof case cover the single surface of a Moebius band.

### 3.3 Symmetric cases.

According to references (McCarthy *et al.* 2011, Mauroy *et al.* 2016) there are twenty-seven possible cases that describe the motion of a four-bar linkage, and among them, for nineteen cases the Grashof condition is accomplished. Specifically, the nineteen cases are those for which at least one of the three variables $r$, $u$, $v$ are zero. Among them, for twelve cases only one of the variables ($r$, $u$, $v$) is zero, for six cases two of the variables ($r$, $u$, $v$) are equal to zero, and there is only one case, where all the three variables are zero.

In this section, we discuss the screens of some of the cases 1 – 19, introduced in Section 3.1. These case, characterized by having at least one of the variables $r$, $u$, $v$ equal to 0, are called symmetric, because of symmetry features presented by the related caustic. It is important to note that for the symmetric cases, the system presents a folded configuration during the rotation of its bars, corresponding to four-bar linkages with $T_1$, $T_2$ or $T_3 = 0$ (McCarthy *et al.* 2011). For the symmetric cases a single revolution along the caustic lead to the initial object.

In Figure 5, we report the screen for a system characterized by the following sides: $a = 100$, $b = 110$, $c = 130$ and $d = 140$; the linkage presents $v = 0$. According to the 6$j$ notation, it is written as follows:

$$\begin{Bmatrix} a & b & x \\ c & d & y \end{Bmatrix} = \begin{Bmatrix} 100 & 110 & X \\ 130 & 140 & Y \end{Bmatrix}. \qquad (36)$$

The linkage is classified as case 8 (see Table I). The diagonals vary within the ranges $10 \leq x \leq 210$ and $40 \leq y \leq 240$. The N and W gates coincides and are located in the upper left corner. This is a particular case of Grashof, for which the summation of the longest and shortest bars are equal to the summation of the intermediate ones, being all the bars of different length. In correspondence of the coalescence the bars lie on the same line, with $a$ and $d$ (in sequence) overlapped to $b$ and $c$ (also displaced in sequence). The passage to the gate E, a triangle with $a$ and $b$ aligned, occurs through a convex quadrilateral, and the passage to the gate S, with $a$ and $d$ overlapped, through a concave quadrilateral. Finally, the passage between S and the coalescence NW involves a biconcave quadrilateral.

In Figure 6, we show the screen for $a = 100$, $b = 140$, $c = 130$ and $d = 110$, with $u = 0$:

$$\begin{Bmatrix} a & b & x \\ c & d & y \end{Bmatrix} = \begin{Bmatrix} 100 & 140 & X \\ 130 & 110 & Y \end{Bmatrix}. \qquad (37)$$

There is a coalescence of the caustic at SE. It corresponds to the case 3 (see Table I), with input and output links of crank type. The ranges of the diagonals are $40 \leq x \leq 240$ and $10 \leq y \leq 210$. The configuration of the four bars is similar to that of the previous case, but the coalescence at SE indicates the folded system involves $a$ and $b$ aligned, and overlapped with $b$ and $c$.

In Figure 7, the screen for $a = 100$, $b = 130$, $c = 140$ and $d = 110$, with $r = 0$ is reported:

$$\begin{Bmatrix} a & b & x \\ c & d & y \end{Bmatrix} = \begin{Bmatrix} 100 & 130 & X \\ 140 & 110 & Y \end{Bmatrix}. \qquad (38)$$

This *caustic* presents coalescence at SW. It corresponds to the case 4 (see Table I), where both input and output links are of crank type. The ranges of the diagonals are $30 \le x \le 230$ and $10 \le y \le 210$. The coalescence at SW characterizes a completely folded system, where the four bars are overlapped. The quadrilateral involved are a convex, at NE, and a concave at NW and SE. The gate N is represented by a triangle, with the bars *a* and *d* aligned, while the gate E is given by a triangle with the bars *a* and *b* aligned.

In Figure 8, we plot the screen for $a = 100$, $b = 200$, $c = 100$ and $d = 200$, corresponding to the case 16 (Table I)

$$\begin{Bmatrix} a & b & x \\ c & d & y \end{Bmatrix} = \begin{Bmatrix} 100 & 200 & x \\ 100 & 200 & y \end{Bmatrix}. \qquad (39)$$

This case presents a higher Regge symmetry, with $u = v = 0$, it implies that the folding of the system occurs in two different configurations. It presents the *Piero line* with a double coalescence at NW and SE gates. The range of the diagonals are $100 \le x \le 300$ and $100 \le y \le 300$. A linkage with these characteristics presents the four bars aligned in correspondence of the two coalescences. More precisely, *a* and *b* lie on the same line and are overlapped to *c* and *d* at SE, while *a* and *d* lie on the same line and are overlapped with *b* and *c* at NW. A convex quadrilateral characterizes NE, while a biconcave quadrilateral characterizes SW.

In Figure 9, we report the case for $a = 100$, $b = 110$, $c = 100$ and $d = 110$, corresponding to the case 16 (Table I).

$$\begin{Bmatrix} a & b & x \\ c & d & y \end{Bmatrix} = \begin{Bmatrix} 100 & 110 & x \\ 100 & 110 & y \end{Bmatrix}. \qquad (40)$$

As seen in the previous figure, $u = v = 0$. It presents Piero's line and coalescences are at NW and SE gates. The ranges of the diagonals are $10 \le x \le 210$ and $10 \le y \le 210$. We note that *r* qualitatively indicates a change in shape of the caustic and the *ridges*. The linkage is substantially similar to that reported for the previous case, although the bars *a* and *d* and the bars *b* and *c* are overlapped between SW and SE, as well as the bars *c* and *d* and the bars *a* and *b* are overlapped between NW and SW, giving a configuration similar to the hands of the clock.

In Figure 10, the case for $a = 100$, $b = 100$, $c = 1000$ and $d = 1000$ is illustrated and corresponds to the case 17 (Table I), with $r = v = 0$ (the system folds in two different configurations)

$$\begin{Bmatrix} a & b & x \\ c & d & y \end{Bmatrix} = \begin{Bmatrix} 100 & 100 & x \\ 1000 & 1000 & y \end{Bmatrix}. \qquad (41)$$

The ranges of *x* and *y* are $0 \le x \le 200$ and $900 \le y \le 1100$. The *caustic* presents a coalescence at NW and SW. At NW linkage presents the bars *a* and *d* aligned and overlapped to the bars *b* and *c*, while at SW the system is completely folded, with the four bars superimposed. At E, *a* and *b* are aligned, forming a triangle, at NE the configuration is that of a convex quadrilateral, while SE is characterized by a concave quadrilateral.

Figure 11 reports the case for $a = 100$, $b = 100$, $c = 100$ and $d = 100$, corresponding to the case 19

$$\begin{Bmatrix} a & b & x \\ c & d & y \end{Bmatrix} = \begin{Bmatrix} 100 & 100 & x \\ 100 & 100 & y \end{Bmatrix} \tag{42}$$

It represents the case with the highest symmetry, with $r = u = v = 0$. Under these conditions, the system folds in three different configurations and the input and output links rotate completely. The ranges of the diagonals are $0 \leq x \leq 200$ and $0 \leq y \leq 200$. The screen presents the Piero's line and coalescence at NW and SE. The coalescence point at NW is characterized by *a* and *d* aligned and overlapped with *b* and *c*, while at SE *a* and *b* are aligned and overlapped with *c* and *d*. The intermediate point at NE is represented by a convex quadrilateral.

## 4. Concluding remarks.

In this work, we have related the tools developed for the theory of angular momentum in quantum mechanics and the four-bar linkage. Specifically, we have investigated relations between the Regge symmetries and the classification of the four-bars mechanism, especially for what concern the Grashof condition. The screen representation has been adopted for the study of some significant case of four-bar linkage, that have been classified in symmetric and general cases. The symmetric cases are those for which at least one of the Regge variables are equal to 0. We have observed that systems characterized by one of the Regge variables equal to 0, present a folded configuration, during the rotation of the bars. Systems characterized by two Regge variables equal to 0, fold in correspondence of two configurations. Finally, systems with all the three Regge variables equal to 0 present folding correspondent to three different configurations. The screens present coalescence of a couple or both couple of gates, for all the linkages the Grashof condition is verified and systems are invariant under Regge symmetry operations. Regarding the general case, it is characterized by having all the Regge variables different from 0, the corresponding screen presents separated gates and the system is not invariant under Regge symmetry operation. For this latter aspect, we have observed that the Regge conjugate has an opposite Grashof behavior, with respect to the starting system. It has been observed that for the Grashof case, clockwise or anticlockwise bar rotations generate two non-connected sequences of enantiomers that from a topological point of view can be seen as the two non-connected faces of a band. For the non-Grashof case, movements originate both enantiomeric configurations through a 4 $\pi$-cycle, showing analogies in topology with the Moebius band.

Future developments will concern the extension of the present work to the most recent application of the four-bar linkage. Some important examples concern the evolutionary trend in the biomechanics of many species (Muller 1996,

Alfaro *et al.* 2004, Lungu *et al.* 2015), often applied to the robotics and engineering, like the kinematics and anatomy of the human knee (Bapat *et al.* 2017, Bulea *et al.* 2012, Date *et al.* 2016) and the morphologic characteristics of elephants that allow them to sustain the huge amount of body weight (Weissengruber *et al.* 2006). Other significative examples are found in spearing appendages of mantis shrimps (Patek *et al.* 2007, Hu *et al.* 2017) and in the operculus of fishes, a protective covering for the gills (Konow *et al.* 2008) and in the prey-processing mechanism of salmon-type fishes. Vertebrates skull in general show a four-bar linkage mechanism, (see the example of birds, fishes and lizards) this model permits to study the relation between the functionality and the structure of these systems and their evolution (Olsen *et al.* 2016). The feeding system of sea-horses, composed by a quick elevation of the head followed by a rotation, is also modelled by a four-bar mechanism (Roos *et al.* 2009). A similar mechanism is also found in the kinematic of insects' wings along the translational and rotational degrees of freedom (Pranay *et al.* 2012) and bacteria (Hernández-Ocaña *et al.* 2016). It is noteworthy also the application of four-bar mechanisms to engineering based on the emulation of the motion of animals, *e. g.* amphibian robots that mimic the skill of the basiliscus lizard to run on the water surface and walk on the ground (Rao *et al.* 2013). For possible involvements in other areas of chemistry see also (Balzani *et al.* 2017, Carrà 2017, Freund 2017, Zecchina *et al.* 2017).

**APPENDIX**

**A.1. Symmetries of 6j symbols and the Regge symmetries.**

Similarly to the Wigner-Racah-Regge approach for the 6*j* symbol, we arrange the six distances as follows,

$$\begin{Bmatrix} a & b & x \\ c & d & y \end{Bmatrix}, \quad \text{(A1)}$$

where the third column corresponds to the diagonals. According to the invariance under permutation of the three columns, other choices of diagonals are possible:

$$\begin{Bmatrix} a & b & x \\ c & d & y \end{Bmatrix} = \begin{Bmatrix} a & x & b \\ c & y & d \end{Bmatrix} = \dots \quad \text{(A2)}$$

in this way, there are 3! = 6 identical symbols. The choice of the diagonal is useful to monitor specific properties of the system (Aquilanti *et al.* 2017c).

The first row of the 6*j* symbol represents a triangular face of the tetrahedron into a vertex, called triad, while the second row corresponds to the convergence of three edges of the tetrahedron. The tetrahedron is composed by four triads: $\{a, b, x\}, \{c, b, y\}, \{a, d, y\}, \{c, d, x\}$. The invariance of the 6*j* symbol under the interchange of upper and lower arguments of any two columns, generated by the combined invariance with respect to permutations of the four triads and the triangles, gives for example the following relations:

$$\begin{Bmatrix} a & b & x \\ c & d & y \end{Bmatrix} = \begin{Bmatrix} c & d & x \\ a & b & y \end{Bmatrix} = \dots \quad (A3)$$

(four ways). In total, there are twenty-four symmetries.

Each symbol reported above has in addition six replicas that Regge discovered and presented in two famous letters (Regge 1958, 1959). Here, we report in full the thirteen lines of the text that Tullio Regge wrote in his second paper:

"we have shown in a previous letter that the true symmetry of Clebsch-Gordan coefficients is much higher than before believed. A similar result has been now obtained for Racah's coefficients. Although no direct connection has been established between these wider symmetries it seems very probable that it will be found in the future. We shall merely state here the results which can be checked very easily with the help of the well known Racah's formula."

Continues with the symmetry relations:

"From the usual tetrahedral symmetry group of the $6j$ we know already that: $\begin{Bmatrix} a & b & c \\ d & e & f \end{Bmatrix} = \begin{Bmatrix} b & a & c \\ e & d & f \end{Bmatrix} = \begin{Bmatrix} a & e & f \\ d & b & c \end{Bmatrix}$ $= \begin{Bmatrix} c & e & d \\ f & b & a \end{Bmatrix} =$ etc.

Our results can be put into the following form:

$$\begin{Bmatrix} a & b & c \\ d & e & f \end{Bmatrix} = \begin{Bmatrix} a & (b+e+c-f)/2 & (b+c+f-e)/2 \\ d & (b+e+f-c)/2 & (c+e+f-b)/2 \end{Bmatrix} = etc."$$

The paper concludes:

"Only the first of these symmetries is essentially new, the others can be obtained from it and the permutational symmetries. We see therefore that there are 144 identical Racah's coefficients. These new symmetries should reduce by a factor 6 the space required for the tabulation of the $6j$. It should be pointed out that this wider 144-group is isomorphic to the direct product of the permutation groups of three and four objects."

Letters (Regge 1958) and (Regge 1959), and the reprinted (Biedenharn *et al.* 1965) are a collection of articles relevant to quantum angular momentum theory, cited by Bargmann 1962 as "Regge's intriguing discovery of unexpected symmetry of $3j$ and $6j$ symbols".

**Acknowledgments.** The authors gratefully acknowledge the Italian Ministry for Education, University and Research (MIUR) for financial support through SIR 2014 (Scientific Independence of Young Researchers), award number: RBSI14U3VF. Robenilson Ferreira is grateful to Brazilian CAPES for a sandwich doctoral (PDSE88881.134388/2016-01) fellowship to the Perugia University.

**Figure 1.**

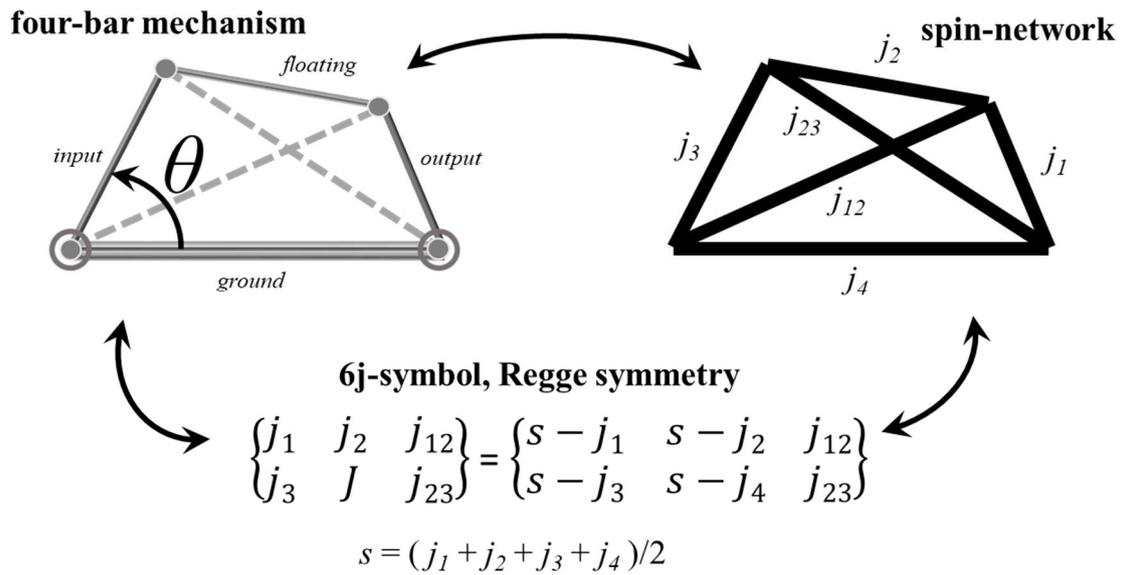

**four-bar mechanism**        **spin-network**

*floating*

*input*    $\theta$    *output*

*ground*

$j_2$, $j_{23}$, $j_3$, $j_1$, $j_{12}$, $j_4$

**6j-symbol, Regge symmetry**

$$\begin{Bmatrix} j_1 & j_2 & j_{12} \\ j_3 & J & j_{23} \end{Bmatrix} = \begin{Bmatrix} s-j_1 & s-j_2 & j_{12} \\ s-j_3 & s-j_4 & j_{23} \end{Bmatrix}$$

$$s = (j_1 + j_2 + j_3 + j_4)/2$$

**Figure 2.**

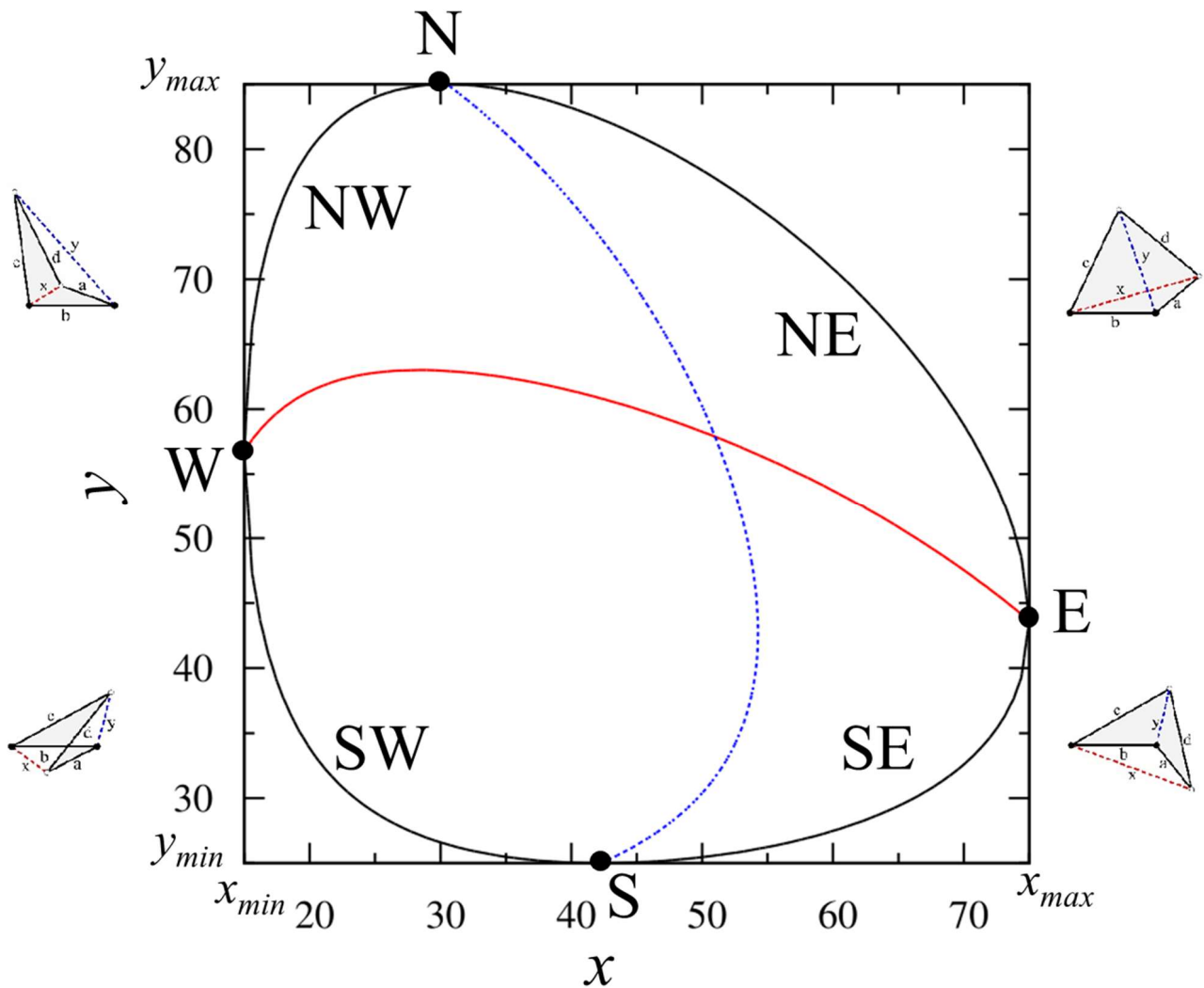

**Figure 3.**

Grashof Case

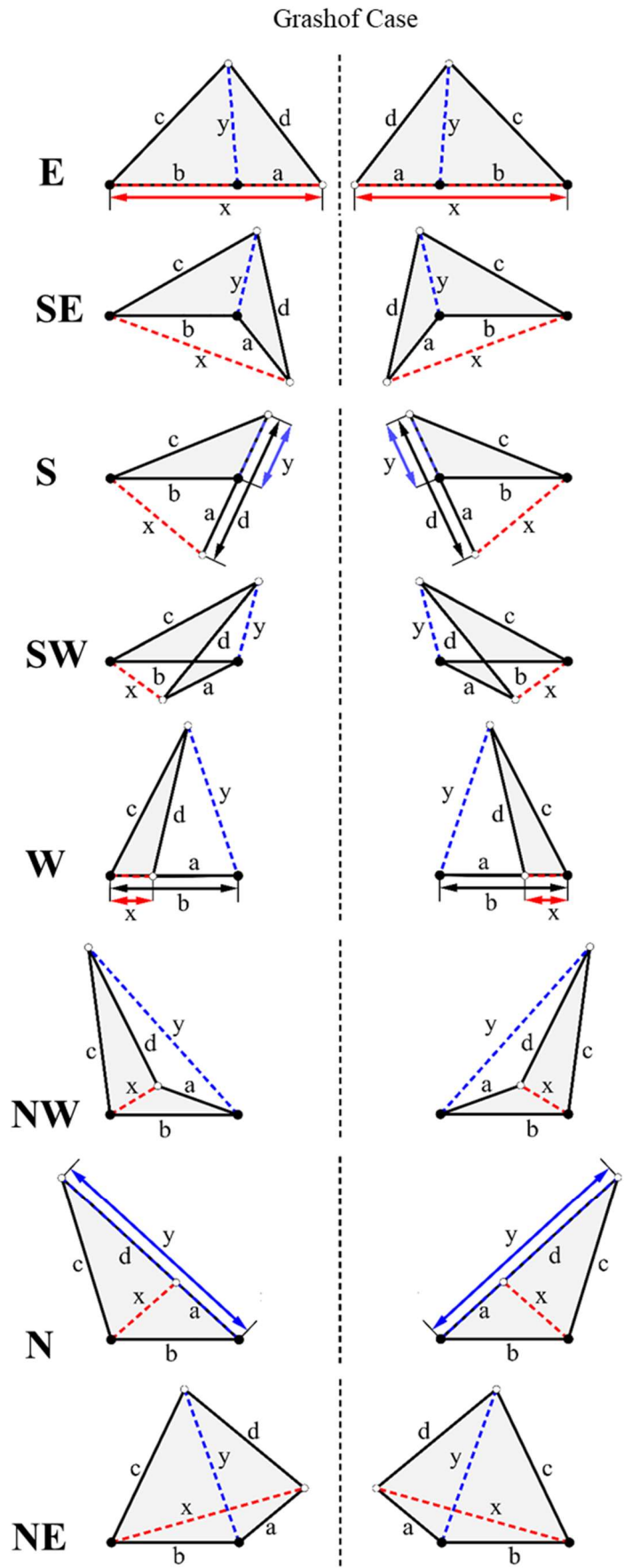

**Figure 4.**

Non-Grashof Case

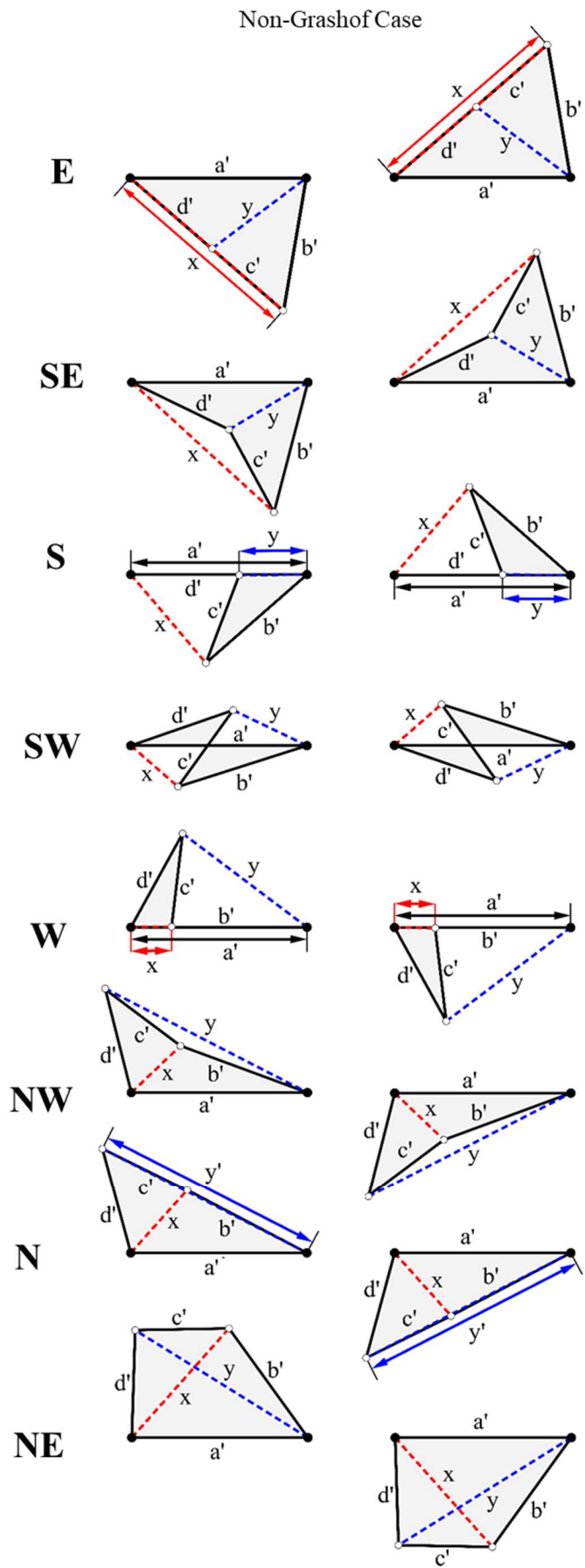

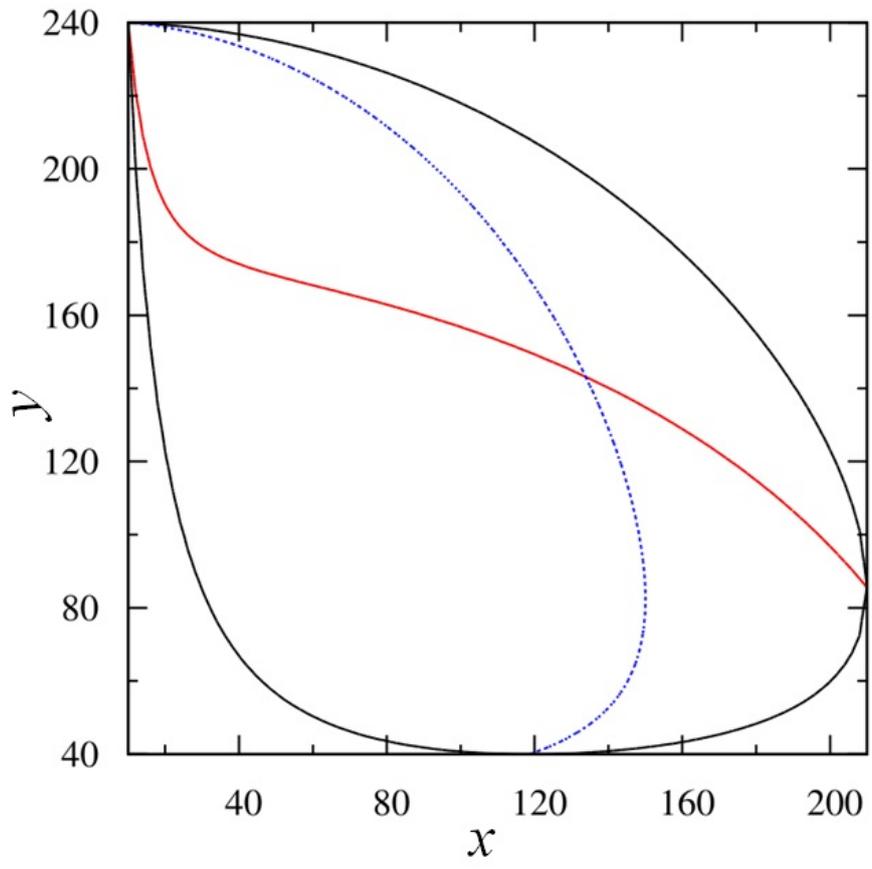

**Figure 6.**

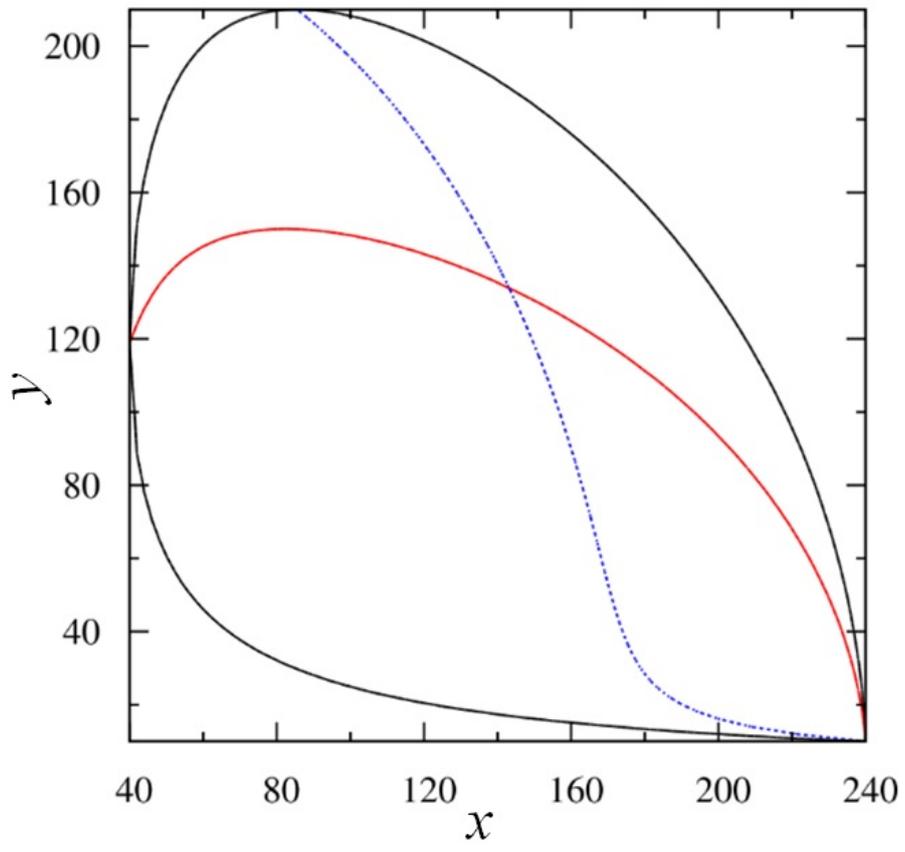

**Figure 7.**

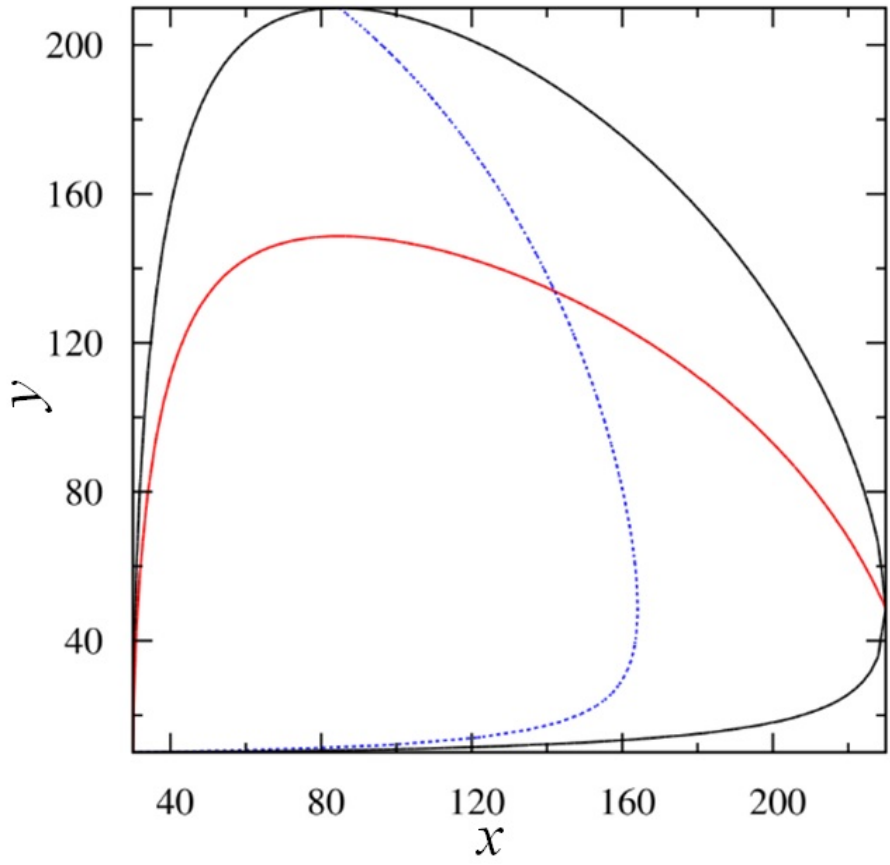

**Figure 8.**

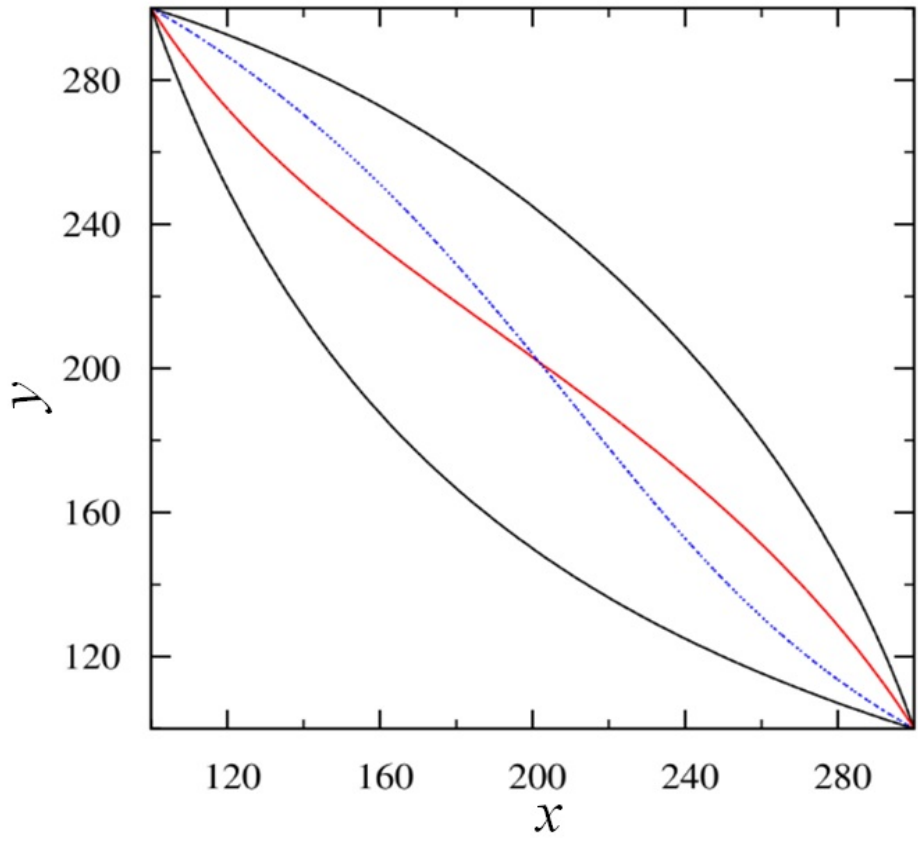

**Figure 9.**

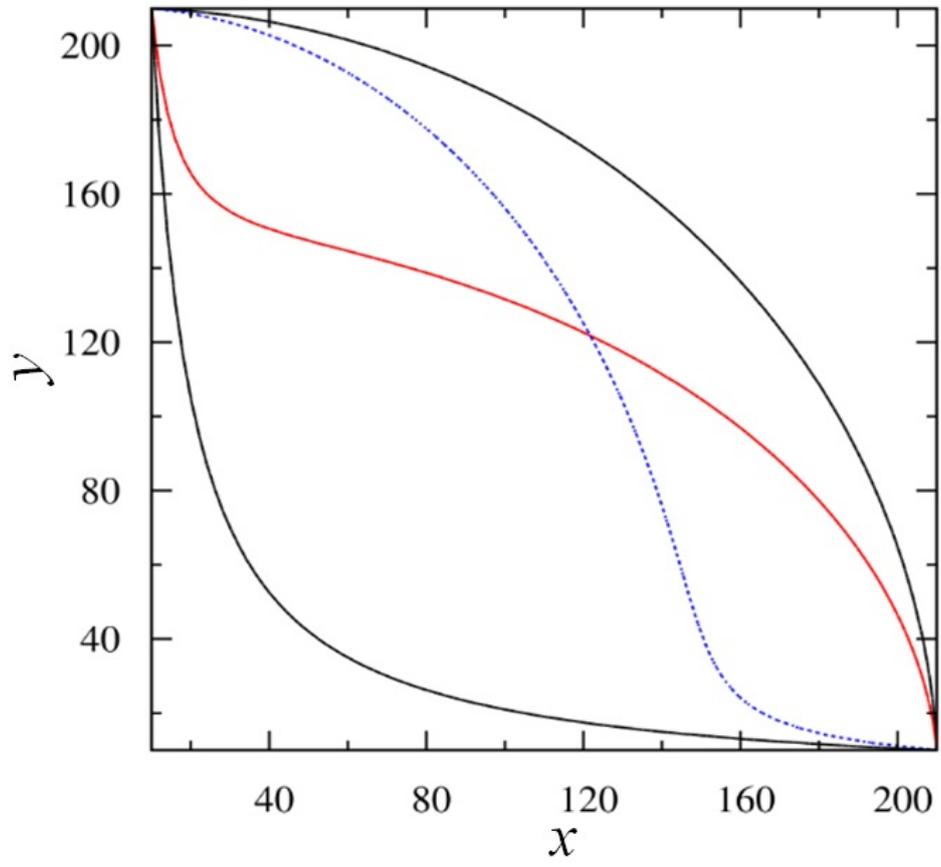

**Figure 10.**

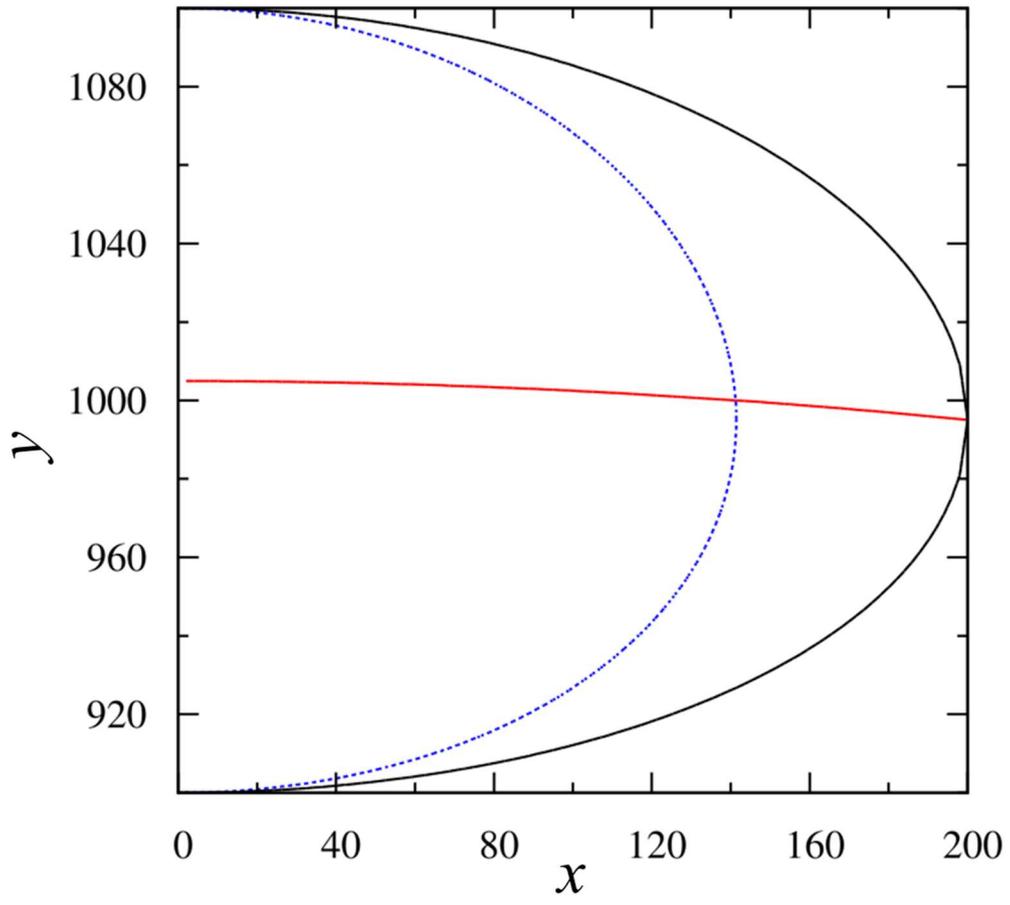

**Figure 11.**

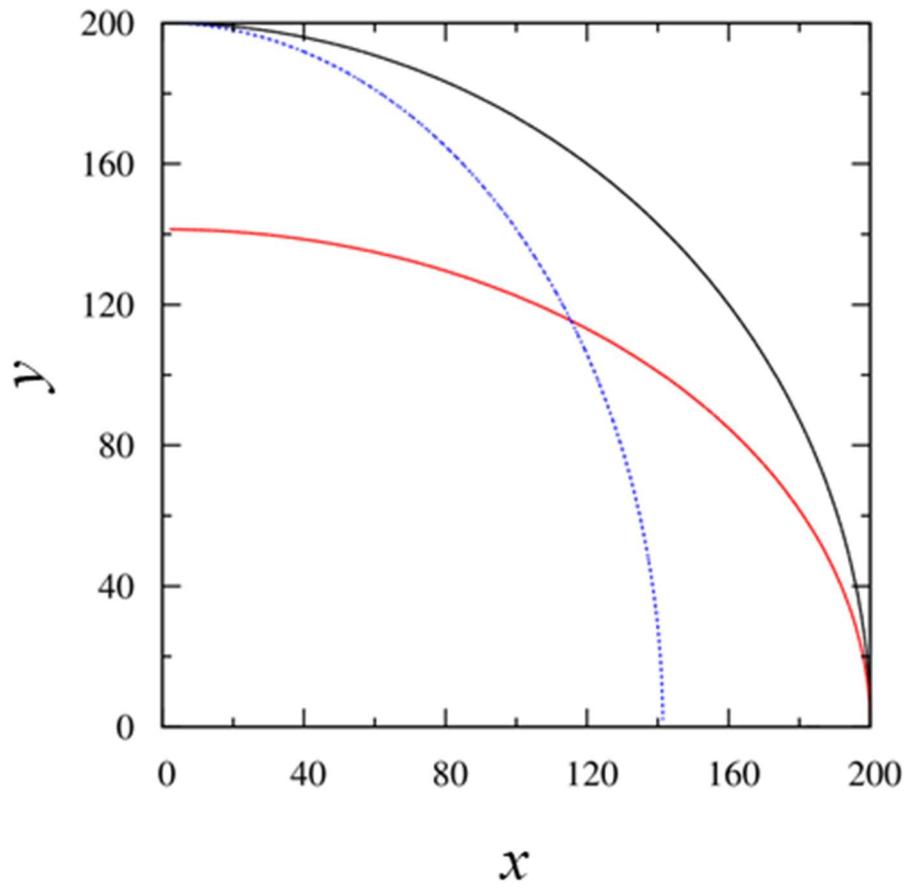

**Captions.**

**Figure 1.** Nomenclature for the four-bar mechanism, spin-network as a complete quadrangle, quantum angular momentum coupling, 6j-symbol and Regge symmetry.

**Figure 2.** The screen of a quadrangle of sides $a = 30$, $b = 45$, $c = 60$, $d = 55$ (for the corresponding 6j symbol screen representation, see Fig. 1 in Ragni *et al.* 2013). The diagram reports the four gates E, S, W, N and four further points SE, SW, NW, NE, the corresponding quadrangle, for these latter, is concave, biconcave, concave and convex, respectively.

**Figure 3.** Illustration of the four-bar mechanism related to the caustic in Figure 2. For this system, the Grashof condition is verified. We report the four sides $a$, $b$, $c$ and $d$, and the diagonals $x$ and y. The side $b$ is the "ground link" and $a$, the "input" or "output link", acts as a "crank" when rotates clockwise (see the left column) or anticlockwise (see the right column). We report the configurations that the system assumes in correspondence of the gates N, S, W, E and the intermediate points NE, SE, NW, SW. For each of this point, there are a couple of mirror images originated independently by the clockwise and anticlockwise rotations. The two sequence of mirror images behave as independent "enantiomers", adopting the nomenclature from molecular chirality, covering the non-connected surfaces of a two-sided band.

**Figure 4.** We report the Grashof conjugate of the object in Figure 3, which bars have lengths $a'$, $b'$, $c'$, $d' = 65, 50, 35, 40$, while $x$ and $y$ are the diagonals. In this case the Grashof condition is not verified. In this illustration, $a'$ is fixed (the "ground link") $b'$ behave as input or output link. From the E configuration $b'$ rotates toward the ground link describing all the configurations until reaching an inversion point at NE. The bar $b'$ does not behave as a crank, as typical of a non-Grashof case. Thus, the system spans all the configurations reported in the right side of the column, that represents a sequence of mirror images, enantiomers, of the configurations in the left side of the column. All the configurations are covered by a twice full rotation around the caustic (Figure 2). The reader can appreciate here the closed sequence obtained from a topological point of view as the covering of the single surface of a Moebius band.

**Figure 5.** Screen of a system characterized by $a = 100$; $b = 110$; $c = 130$; $d = 140$ (for the corresponding 6j symbol screen representation, see Fig. 5a in Ragni *et al.* 2013 and Fig. 1b in Bitencourt *et al.* 2012). It corresponds to the case of symmetry $r = -10$; $u = -30$; $v = 0$; (case 8, see Table I-b). The N and W gates coincide.

**Figure 6.** The screen of the system of sides $a = 100$; $b = 140$; $c = 130$; $d = 110$ (for the corresponding 6j symbol screen representation, see Fig5b in Ragni *et al.* 2013 and Fig 1c in Bitencourt *et al.* 2012 ) is reported. The symmetry parameters are $r = -10$; $u = 0$; $v = 30$ (case 3, see Table I-b) With respect to the system reported in a), there is an inversion in the shape of the curve, with the S and E gates coinciding.

**Figure 7.** The screen is related to the system $a = 100$; $b = 130$; $c = 140$; $d = 110$ (for the corresponding 6j symbol screen representation, see Fig. 6a in Ragni *et al.*2013 and Fig 1d in Bitencourt *et al.* 2012) and symmetry parameters $r = 0$; $u = -10$; $v = 30$ (case 4, see Table I-b). There is coalescence between the South and West gates.

**Figure 8.** Screen of the system $a = 100$; $b = 200$; $c = 100$; $d = 200$ (for the corresponding 6j symbol screen representation, see Fig. 8a in Ragni *et al.*2013 and Fig 4a in Bitencourt *et al.* 2012). The symmetry parameters are $r = -100$ and $u = v = 0$ (case 16, see Table I-c). There is coalescence between the gates North and West and the gates South and East.

**Figure 9.** In this panel, we report the screen for the system $a = 100$; $b = 110$; $c = 100$; $d = 110$ (for the corresponding 6j symbol screen representation, see Fig. 8b in Ragni *et al.* 2013 and Fig 4b in Bitencourt *et al.* 2012). The symmetry parameters are $r = -10$ and $u = v = 0$ (case 16, see Table I-c).

**Figure 10.** Screen for the system $a = 100$; $b = 100$; $c = 1000$; $d = 1000$ (for the corresponding 6j symbol screen representation, see Fig. 9a in Ragni et al 2013 and Fig 5 in Bitencourt *et al.* 2012). In this case, the symmetry parameters are $r = 0$; $u = -900$; $v = 0$ (case 17, see Table I-c).

**Figure 11.** Screen of the system $a = b = c = d = 100$ (for the corresponding 6j symbol screen representation, see Fig. 9b in Ragni *et al.* 2013 and Fig.6 in Bitencourt *et al.* 2012), a quadrilateral with the four sides of identical length. The symmetry parameters are $r = u = v = 0$ (case 19, see Table I-c).

**Table I-a. Classification of movements of input and output links according to r, u, v variables: A general cases, r, u, v ≠ 0**

| Notation | Variables | | | Movements | |
|:---:|:---:|:---:|:---:|:---:|:---:|
| | r | u | v | *Input link* | Output link |
| I | - | - | + | crank | crank |
| I' | + | + | - | 0-rocker | π -rocker |
| II | + | - | - | rocker | crank |
| II' | - | + | + | π -rocker | π -rocker |
| III | - | + | - | rocker | rocker |
| III' | + | - | + | π -rocker | 0-rocker |
| IV | + | + | + | Crank | rocker |
| IV' | - | - | - | 0-rocker | 0-rocker |

**\* Unprimed and primed Roman numerals distinguish Regge Conjugates. Unprimed ones are Grashof, i.e. perform crank rotations (in case III the crank is the floating link)**

**Table I-b. Classification of movements of the input and output links. Case where only one of *r, u, v* variables is zero.**

| Notation | r | u | v | Input link | Output link |
|----------|---|---|---|------------|-------------|
| 1 | 0 | + | + | Crank | π -rocker |
| 2 | + | 0 | + | crank | 0-rocker |
| 3 | - | 0 | + | crank | crank |
| 4 | 0 | - | + | crank | crank |
| 5 | + | + | 0 | crank | π -rocker |
| 6 | - | + | 0 | π -rocker | π -rocker |
| 7 | + | - | 0 | π -rocker | crank |
| 8 | - | - | 0 | crank | crank |
| 9 | 0 | + | - | 0-rocker | π -rocker |
| 10 | + | 0 | - | 0-rocker | crank |
| 11 | - | 0 | - | 0-rocker | 0-rocker |
| 12 | 0 | - | - | 0-rocker | crank |

**Table I-c. Case where at least two of the r, u, v variable are zero.**

| Notation | r | u | V | *Input link* | Output link |
|----------|---|---|---|------------|-------------|
| 13 | 0 | 0 | + | crank | crank |
| 14 | 0 | + | 0 | crank | π -rocker |
| 15 | + | 0 | 0 | crank | crank |
| 16 | - | 0 | 0 | crank | crank |
| 17 | 0 | - | 0 | crank | crank |
| 18 | 0 | 0 | - | 0-rocker | crank |
| 19 | 0 | 0 | 0 | crank | crank |

(\* Roman numerals refer to Regge's symmetric in the case where *r, u, v* are different from zero, while the Arabic numbers refer to the other 19 symmetry cases where *r, u, v* = 0)

**Table 4. Transformation relation between 6j case and Regge's symmetry**

| Case 1 | Regge 's Case1 | Case 2 | Regge 's Case 2 | Case 3 | Regge 's Case 3 | Case 4 | Regge 's Case 4 |
|---|---|---|---|---|---|---|---|
| $\begin{pmatrix} 2r \\ 2u \\ 2v \end{pmatrix}$ | $\begin{pmatrix} -2r \\ -2u \\ -2v \end{pmatrix}$ | $\begin{pmatrix} -2r \\ 2u \\ -2v \end{pmatrix}$ | $\begin{pmatrix} 2r \\ -2u \\ 2v \end{pmatrix}$ | $\begin{pmatrix} 2r \\ -2u \\ -2v \end{pmatrix}$ | $\begin{pmatrix} -2r \\ 2u \\ 2v \end{pmatrix}$ | $\begin{pmatrix} -2r \\ -2u \\ 2v \end{pmatrix}$ | $\begin{pmatrix} 2r \\ 2u \\ -2v \end{pmatrix}$ |
| $\begin{pmatrix} -60 \\ -160 \\ 60 \end{pmatrix}$ | $\begin{pmatrix} 60 \\ 160 \\ -60 \end{pmatrix}$ | $\begin{pmatrix} 60 \\ -160 \\ -60 \end{pmatrix}$ | $\begin{pmatrix} -60 \\ 160 \\ 60 \end{pmatrix}$ | $\begin{pmatrix} -60 \\ 160 \\ -60 \end{pmatrix}$ | $\begin{pmatrix} 60 \\ -160 \\ 60 \end{pmatrix}$ | $\begin{pmatrix} 60 \\ 160 \\ 60 \end{pmatrix}$ | $\begin{pmatrix} -60 \\ -160 \\ -60 \end{pmatrix}$ |
| $\begin{pmatrix} -10 \\ -40 \\ 20 \end{pmatrix}$ | $\begin{pmatrix} 10 \\ 40 \\ -20 \end{pmatrix}$ | $\begin{pmatrix} 10 \\ -40 \\ -20 \end{pmatrix}$ | $\begin{pmatrix} -10 \\ 40 \\ 20 \end{pmatrix}$ | $\begin{pmatrix} -10 \\ 40 \\ -20 \end{pmatrix}$ | $\begin{pmatrix} 10 \\ -40 \\ 20 \end{pmatrix}$ | $\begin{pmatrix} 10 \\ 40 \\ 20 \end{pmatrix}$ | $\begin{pmatrix} -10 \\ -40 \\ -20 \end{pmatrix}$ |
| Transformation | | | | | | | |
| $(-1)\begin{pmatrix} r \\ u \\ v \end{pmatrix} = \begin{pmatrix} r' \\ u' \\ v' \end{pmatrix}$ | | $(-1)\begin{pmatrix} -r \\ u \\ -v \end{pmatrix} = \begin{pmatrix} r' \\ -u' \\ v' \end{pmatrix}$ | | $(-1)\begin{pmatrix} r \\ -u \\ -v \end{pmatrix} = \begin{pmatrix} -r' \\ u' \\ v' \end{pmatrix}$ | | $(-1)\begin{pmatrix} -r \\ -u \\ v \end{pmatrix} = \begin{pmatrix} r' \\ u' \\ -v' \end{pmatrix}$ | |